# Relationship between brain injury criteria and brain strain across different types of head impacts can be different.

**Xianghao Zhan***: Department of Bioengineering, Stanford University, Stanford, CA, 94305, USA; xzhan96@stanford.edu

**Yiheng Li**: Department of Biomedical Informatics, Stanford University, Stanford, CA, 94305, USA; yyhhli@stanford.edu

**Yuzhe Liu**: Department of Bioengineering, Stanford University, Stanford, CA, 94305, USA; yuzheliu@stanford.edu

**August G. Domel**: Department of Bioengineering, Stanford University, Stanford, CA, 94305, USA; augustdomel@gmail.com

**Hossein Vahid Alizadeh**: Department of Bioengineering, Stanford University, Stanford, CA, 94305, USA; hva@stanford.edu

**Samuel J. Raymond**: Department of Bioengineering, Stanford University, Stanford, CA, 94305, USA; sjray@stanford.edu

**Jesse Ruan**: Ford Motor Company, 3001 Miller Rd, Dearborn, MI 48120, USA; jruan@ford.com

**Saeed Barbat**: Ford Motor Company, 3001 Miller Rd, Dearborn, MI 48120, USA; sbarbat@ford.com

**Stephen Tiernan:** Technological University Dublin, Dublin, Ireland; stephen.tiernan@tudublin.ie

**Olivier Gevaert**: Department of Biomedical Informatics, Stanford University, Stanford, CA, 94305, USA; ogevaert@stanford.edu

**Michael Zeineh**: Department of Radiology, Stanford University, Stanford, CA, 94305, USA; mzeineh@stanford.edu

**Gerald Grant**: Department of Neurosurgery, Stanford University, Stanford, CA, 94305, USA; ggrant2@stanford.edu

**David B. Camarillo**: Department of Bioengineering, Stanford University, Stanford, CA, 94305, USA; dcamarillo@stanford.edu

Xianghao Zhan, Yiheng Li and Yuzhe Liu contributed equally to this work.
*Corresponding author: Xianghao Zhan

**Abstract**

Multiple brain injury criteria (BIC) are developed to quickly quantify brain injury risks after head impacts. These BIC originated from different types of head impacts (e.g., sports and car crashes) are widely used in risk evaluation. However, the accuracy of using the BIC on brain injury risk estimation across different types of head impacts has not been evaluated. Physiologically, the brain strain is often considered the key parameter of brain injury. To evaluate the BIC's risk estimation accuracy across five datasets comprising different head impact types, linear regression was used to model 95% maximum principal strain, 95% maximum principal strain at corpus callosum, and cumulative strain damage (15%) on each of 18 BIC respectively. The results show significant difference in the relationship between BIC and brain strain across datasets, indicating the same BIC value may suggest different brain strain in different head impact types. The accuracy of brain strain regression is generally decreasing if the BIC regression models are fit on a dataset with a different type of head impact rather than on the dataset with the same type. Given this finding, this study raises concerns for applying BIC to estimate the brain injury risks for head impacts different from the head impacts on which the BIC was developed.

**1. Introduction**

Traumatic brain injury (TBI), frequently caused by head impacts in sports, traffic accidents and unintentional falls, has become a global health challenge and affects over 50 million children and adults worldwide (1). Additionally, mild traumatic brain injury (mTBI) is associated with long-term cognitive and emotional sequelae (2), cerebral blood flow alterations (3) and even neurological degenerative diseases (4). If mTBI goes undetected, the accumulation of brain damage causes higher risks of long-term consequences (5), which calls for better monitoring of brain injury after head impacts.

To quickly estimate the brain injury risk of a head impact, multiple brain injury criteria (BIC) have been developed with reduced-order physical models (6–9) and statistical model fitting (10,11). These BIC estimate the risk of brain injury based on the measured kinematics of head movement caused by an impact. Additionally, the peak values of head movement kinematics, such as the linear acceleration at the brain center of gravity, angular velocity and angular acceleration, can also be used as a BIC (8,12,13).

Because the brain can be injured when head movement deforms brain tissue by the inertial force (6–8,14–16), brain strain, particularly the maximum principal strain (MPS), is a key parameter in evaluating the brain injury risk (3,17–19), and has been widely used as an indicator for mTBI. To evaluate the accuracy of estimating brain injury risks of the BIC, typically Pearson correlation between the BIC and the brain strain (particularly 95% MPS), and the coefficient of determination ($R^2$) of the linear regression of brain strain on the BIC, have been used by researchers (6–8,20).

Although the BIC are typically designed for specific types of head impacts, many BIC are used regardless of their development background. For instance, head injury criterion (HIC) originated in the dummy test of the automobile crashworthiness and in the field of motor vehicle regulation (21), but is now used in mTBI research (22). However, when cross-used to different types of head impacts, even for the same values of MPS, some BIC give different values. As is shown in **Fig. 1**, the impacts of college football and mixed martial arts show different characteristics in their kinematics and the BIC calculated based on the kinematics (Section 2.2) show clear distinctions. However, these two impacts reach the same peak value of 95% MPS. That different BIC values indicate potentially the same brain strain in different types of head impacts, poses a risk to the general use of these BIC across the mTBI field.

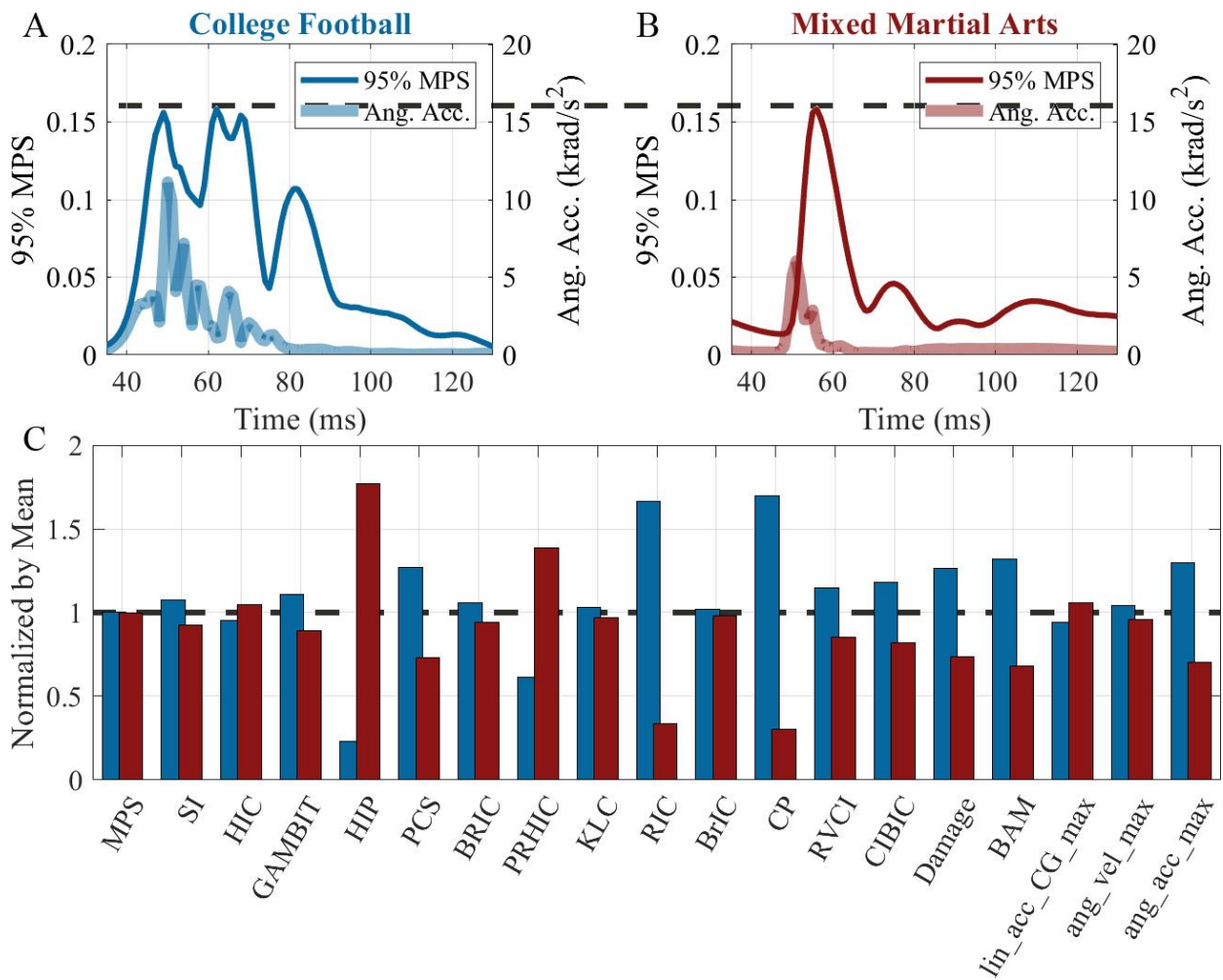

**Figure 1** Comparison between two impacts in college football (A) and in mixed martial arts (B). The traces of 95% MPS (left axis) and angular acceleration (Ang. Acc., right axis) magnitudes are plotted in (A, B), and these two impacts have the same peak value of 95% MPS as shown by the dash line across (A) and (B). The 95% MPS peak value and BIC values are given in (C). The blue and red bars indicate college football and mixed martial arts, respectively. The information of BIC is introduced in Section 2.2.

      To better understand the risk of using a BIC as a general metric, in this work several datasets from various sports and various types of head impacts are used to investigate the cross-usage of BIC. Linear regression models were built with 18 different BIC as predictors for 95% MPS (MPS95), 95% MPS at corpus callosum (MPSCC95) and cumulative strain damage (CSDM ,15%, indicating the volume fraction of brain with MPS exceeding the threshold of 0.15 (23)), to investigate the regression coefficients of the BIC and the brain strain estimation accuracy across datasets. These models were tested against the different datasets in a variety of statistical strategies to best determine the efficacy of using the BIC across datasets.

**2. Methods**
**2.1 Data description and processing**

To evaluate the generalizability of BIC across different types of head impacts, data from six individual sources were used to construct five datasets: 1) the dataset 1 contained 2183 lab impacts, among which 2130 were simulated by a validated finite element analysis (FEA) model of hybrid III anthropomorphic test dummy (ATD) headform impacted by the impactor for football testing (20,24,25). The remaining 53 were reconstructed head impacts in National Football League (26) using hybrid III ATD headform. The data from these two sources were merged because they both represented the responses of hybrid III ATD headform under football-type impacts; 2) the dataset 2 contained 302 on-site college football head impacts collected by the Stanford instrumented mouthguards (27–29); 3) the dataset 3 contained 457 on-site mixed martial arts (MMA) head impacts collected by the Stanford instrumented mouthguard (3,30); 4) the dataset 4 contained 48 head impacts in automobile crashworthiness tests from the National Highway Traffic Safety Administration (NHTSA) (31); 5) the dataset 5 contained 272 numerically reconstructed head impacts in National Association for Stock Car Auto Racing (NASCAR) by hybrid III ATD headform.

Since finite element (FE) modeling is the state-of-the-art biomechanics modeling tool for brain strain in head impacts, the validated KTH model (32) was used to calculate the MPS95, MPSCC95 and CSDM since these are widely used in mTBI research (6,7,23,33,34). The KTH model includes the brain, skull, scalp meninges, cerebrospinal fluid (CSF) and 11 pairs of bridging veins. The modeling was performed using the commercially available FE software LS-DYNA (Livermore, CA, USA). The accuracy in estimating the MPS95/MPSCC95/CSDM was then used to evaluate the accuracy of brain injury risk estimation based on each BIC. The distributions and the summary statistics of the MPS95, MPSCC95 and CSDM in the five datasets are shown in **Fig. 2**.

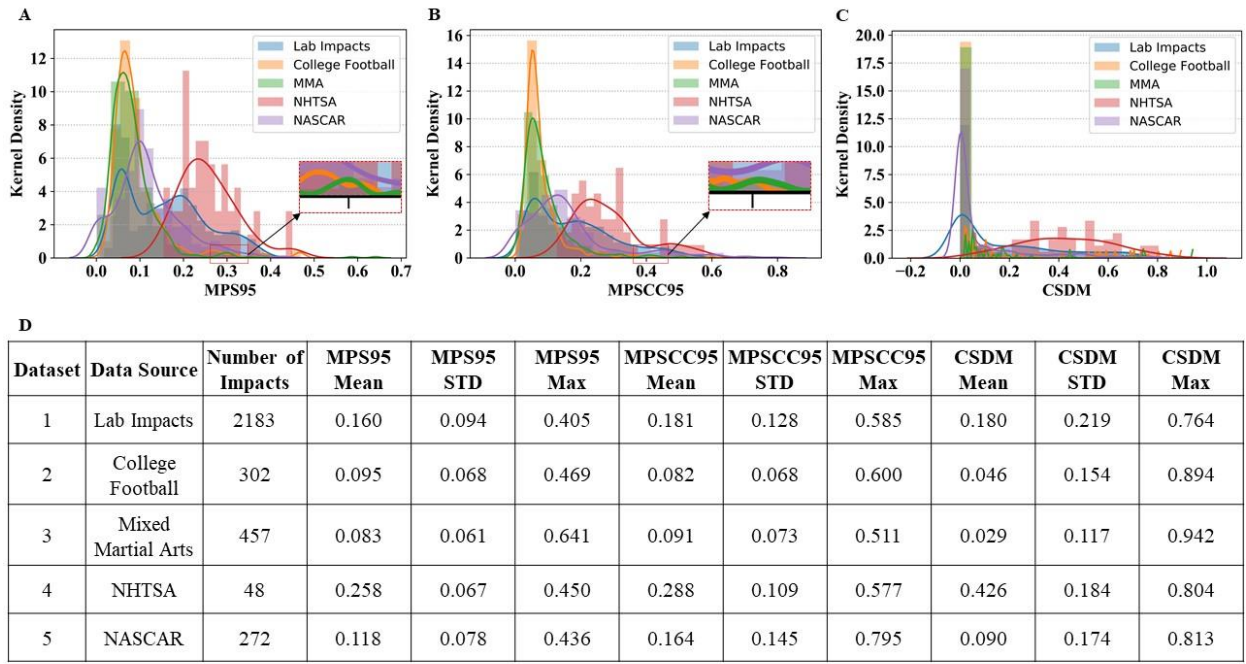

| Dataset | Data Source | Number of Impacts | MPS95 Mean | MPS95 STD | MPS95 Max | MPSCC95 Mean | MPSCC95 STD | MPSCC95 Max | CSDM Mean | CSDM STD | CSDM Max |
|---|---|---|---|---|---|---|---|---|---|---|---|
| 1 | Lab Impacts | 2183 | 0.160 | 0.094 | 0.405 | 0.181 | 0.128 | 0.585 | 0.180 | 0.219 | 0.764 |
| 2 | College Football | 302 | 0.095 | 0.068 | 0.469 | 0.082 | 0.068 | 0.600 | 0.046 | 0.154 | 0.894 |
| 3 | Mixed Martial Arts | 457 | 0.083 | 0.061 | 0.641 | 0.091 | 0.073 | 0.511 | 0.029 | 0.117 | 0.942 |
| 4 | NHTSA | 48 | 0.258 | 0.067 | 0.450 | 0.288 | 0.109 | 0.577 | 0.426 | 0.184 | 0.804 |
| 5 | NASCAR | 272 | 0.118 | 0.078 | 0.436 | 0.164 | 0.145 | 0.795 | 0.090 | 0.174 | 0.813 |

**Figure 2** The normalized distribution and the summary statistics of MPS95, MPSCC95 and CSDM on five datasets. (A) The normalized distribution of MPS95 on five datasets. (B) The normalized distribution of MPSCC95 on five datasets. (C) The normalized distribution of CSDM on five datasets. (D) The summary statistics (mean values, standard deviations (STD) and maximum values) of MPS95/MPSCC95/CSDM on five datasets.

## 2.2 Brain Injury Criteria

In this study, 18 BIC were investigated which have been used in TBI/mTBI research. These are listed in **Table 1** (Lin. Acc.: Linear Acceleration, Ang. Acc.: Angular Acceleration, Ang. Vel.: Angular Velocity). These BIC were based on one or a combination of translational and rotational parameters in the head movement kinematics with experimental data fitting or reduced-order mechanical models (6–8).

The peak values of the magnitudes of the linear acceleration at the brain center of gravity $|a(t)|$ (lin_acc_CG_max) (12), angular velocity $|\omega(t)|$ (ang_vel_max) (13) and angular acceleration $|\alpha(t)|$ (ang_acc_max) (13,20) are the most fundamental BIC definitions. They are calculated by taking the maximum of the magnitude of the respective translational or rotational parameters. In addition to these three BIC, 15 other BIC used in this study are described as follows:

Severity Index (SI), also known as Gadd Severity Index (GSI), was developed by Gadd (35) based on curve fitting the Wayne State Tolerance Curve with skull fracture data in head form simulation. It is calculated based on the following equation, and the integral is calculated from when the signal first exceeds 4g m/s² to when it returns to 4g m/s² after the highest peak:

$$\text{SI} = \int |a(t)|^{2.5} dt \tag{Eqn. 1}$$

Head Injury Criterion (HIC) was developed by Versace et al. (21). It is the most widely used injury criteria developed in the assessment of head injury risks in motor vehicle regulation. HIC is calculated based on the following equation, where $t_1, t_2$ are chosen to maximize the value of HIC. The duration was initially set as 36ms and under the current standard $t_2 - t_1 \leq 15ms$, and the resultant term is expressed as $\text{HIC}_{15}$.

$$\text{HIC} = \max_{t_1,t_2}\left\{\left(\int_{t_1}^{t_2}|a(t)|dt\right)^{2.5}(t_2 - t_1)\right\} \tag{Eqn. 2}$$

Generalized Acceleration Model for Brain Injury Threshold (GAMBIT) was developed by Newman et al. (36). It was designed to combine rotational and translational components of head accelerations, as an analogue in the kinematics to maximum shear stress/strain theory. It is calculated based on the following equation, where the $a_c, \alpha_c$ are the thresholds for the corresponding acceleration and the constants are: $n = m = s = 2, a_c = 250g$ and $\alpha_c = 25000 rad/s^2$.

$$\text{GAMBIT} = \left\{\left[\left(\frac{|a(t)|}{a_c}\right)^n + \left(\frac{|\alpha(t)|}{\alpha_c}\right)^m\right]^{1/s}\right\} \tag{Eqn. 3}$$

Head Impact Power (HIP) was developed by Newman (37). It was based on the hypothesis that the severity of head injury correlated with the head impact power. HIP is calculated based on the following equation, where $m$ denotes the mass and $I_{ii}$ denotes the principal moments of head inertia, and $i$ denotes the components of the acceleration in three spatial directions:

$$\text{HIP} = \max_t\{m\sum a_i(t)\int a_i(t)dt + \sum I_{ii}\alpha_i(t)\int \alpha_i(t)dt\} \tag{Eqn. 4}$$

Principal Component Score (PCS) was developed by Greenwald et al. (11) using football data and principal component analysis (PCA). It is calculated as a linear combination of HIC, SI, maximum magnitude of linear acceleration and maximum magnitude of angular acceleration, while each term of the kinematics is standardized and the coefficients are fitted empirically (11):

$$\text{PCS} = \beta_0 + \beta_1 |a(t)| + \beta_2 \text{ SI} + \beta_3 \text{ HIC} + \beta_4 |\alpha(t)| \tag{Eqn. 5}$$

Kinematic rotational brain injury criterion (BRIC) was developed by Takhounts et al. (38) by adopting the effects of both angular acceleration and angular velocity. The critical values $\omega_{cr}$ and $\alpha_{cr}$ were design variables and decided by risk of diffuse axonal injury (DAI) and the best linear fit between CSDM and the BRIC. Different $\omega_{cr}$ and $\alpha_{cr}$ were given in (38) and the parameters used were obtained by the on-field football data.

$$\text{BRIC} = \frac{|\omega|}{\omega_{cr}} \frac{|\alpha|}{\alpha_{cr}} \qquad \text{(Eqn. 6)}$$

Power Rotation Head Injury Criterion (PRHIC) was developed by Kimpara et al. (39) with modifications on the HIC. It is calculated based on the following equation under the constraint $t_2 - t_1 \leq 36ms$, where $HIP_{rot}$ is the second part of HIP contributed by the rotation:

$$\text{PRHIC} = \left\{ \left( \int_{t_1}^{t_2} |HIP_{rot}(t)| dt \right)^{2.5} (t_2 - t_1) \right\} \qquad \text{(Eqn. 7)}$$

Kleiven's Linear Combination (KLC) was proposed by Kleiven (40) as a brain injury predictor based on a combination of $HIC_{36}$ and the maximum magnitude of angular velocity. It is calculated according to the following formula:

$$\text{KLC} = 0.004718 \max_t |\omega(t)| + 0.000224 HIC_{36} \qquad \text{(Eqn. 8)}$$

Rotational Injury Criterion (RIC) was developed by Kimpara et al. (41) in a similar form to HIC but with the linear acceleration replaced by the angular acceleration. RIC is calculated based on the following equation under the constraint $t_2 - t_1 \leq 36ms$:

$$\text{RIC} = \left\{ \left( \int_{t_1}^{t_2} |\alpha(t)| dt \right)^{2.5} (t_2 - t_1) \right\} \qquad \text{(Eqn. 9)}$$

Brain Injury Criterion (BrIC) was developed by Takhounts et al. (42) based on the assumption that strains calculated by FE modeling can be predicted by angular velocity alone in pendulum and occupant crash tests. It is calculated based on the following formula, where $[\omega_x, \omega_y, \omega_z]$ are the maximum value of the angular velocity in three spatial directions, and $[\omega_{xcr}, \omega_{ycr}, \omega_{zcr}]$ are the corresponding critical values [66.2, 59.1, 44.2] rad/s found by experimental data:

$$\text{BrIC} = \sqrt{\left(\frac{\omega_x}{\omega_{xcr}}\right)^2 + \left(\frac{\omega_y}{\omega_{ycr}}\right)^2 + \left(\frac{\omega_z}{\omega_{zcr}}\right)^2} \qquad \text{(Eqn. 10)}$$

Combined Probability of Concussion (CP) was developed by Rowson et al. (10). It is based on the risk of concussion in the football head impacts with the logistic function. The metric is calculated as the basis of the logistic function in a linear combination of the maximum magnitude of translational and rotational accelerations and the interaction between these two terms, where each of the coefficients is determined by logistic regression:

$$\text{CP} = \beta_0 + \beta_1 |a(t)| + \beta_2 |\alpha(t)| + \beta_3 |a(t)| |\alpha(t)| \qquad \text{(Eqn. 11)}$$

Rotational Velocity Change Index (RVCI) was developed by Yanaoka et al. (43) based on pedestrian impact events. It assumed an analogy between the brain strain and the deformation of a spring-mass model. It is calculated based on the following formula, where $R_i$ are the weighing factors related to each axis, which is determined by FE model, and the duration constraint was chosen to be $t_2 - t_1 \leq 10ms$:

$$\text{RVCI} = \max_{(t_1, t_2)} \sqrt{R_x (\int_{t_1}^{t_2} \alpha_x dt)^2 + R_y (\int_{t_1}^{t_2} \alpha_y dt)^2 + R_z (\int_{t_1}^{t_2} \alpha_z dt)^2} \qquad \text{(Eqn. 12)}$$

Convolution of impulse response for Brain Injury Criterion (CIBIC) was developed by Takahashi et. al. (9). CIBIC used the similar lumped-mass system as Damage (7) but the coupling effects in different directions were not considered. Furthermore, brain strain caused by the impulse with 1ms was used to solve the system, and the displacement, which indicated the 95% MPS, was calculated by the convolution integral.

Damage was developed by Gabler et. al (7) based on a three-degree-of-freedom, 2nd-order lumped-mass system. To include the coupling effects between direction directions, the off-diagonal elements in the stiffness and damping matrices were assumed to be non-zero. The head angular acceleration was used as the input and the displacement of the mass was assumed to indicate the 95% MPS.

Brain Angle Metric (BAM) was developed by Laksari et al. (8) based on a 3 degree-of-freedom lumped parameter brain model and a combined dataset with head impacts from college football, high school football, navy sled tests, etc. BAM is calculated by taking the maximum brain angle $\vec{\theta}_{brain}$ in each direction. It is based on the following equations, where the k and c are the stiffness and damping coefficient of the system and the $\theta_{brain}, \theta_{skull}$ denote the angles of the brain and the skull, with the related model parameters specified (8):

$$I(\ddot{\theta}_{brain} + \ddot{\theta}_{skull}) = -k\theta_{brain} - c\dot{\theta}_{brain} \qquad \text{(Eqn. 13)}$$

**Table 1. The 18 BIC analyzed in this study**

| BIC Name | Equation | Source Impact | Kinematics Included | FE model | Reference |
| --- | --- | --- | --- | --- | --- |
| SI | 1 | Car Crash | Lin. Acc. | - | (35) |
| HIC | 2 | Car Crash | Lin. Acc. | - | (21) |
| GAMBIT | 3 | Car Crash | Lin. Acc., Ang. Acc. | - | (36) |
| HIP | 4 | Simulation | Lin. Acc., Ang. Acc. | - | (37) |
| PCS | 5 | Football | Lin. Acc., Ang. Acc. | - | (11) |
| BRIC | 6 | Football | Ang. Acc., Ang. Vel. | SIMon(44) | (38) |
| PRHIC | 7 | Football | Ang. Acc. | THUMS(45) | (39) |
| KLC | 8 | Football | Lin. Acc., Ang. Acc. | KTH | (40) |
| RIC | 9 | Football | Ang. Acc. | THUMS | (41) |
| BrIC | 10 | Simulation, Football, Car Crash | Ang. Vel. | GHBMC(46) | (42) |
| CP | 11 | Football | Lin. Acc., Ang. Acc. | - | (10) |
| RVCI | 12 | Car Crash | Ang. Vel.* | GHBMC | (43) |
| CIBIC | - | Car Crash | Ang. Acc. | GHBMC | (9) |
| Damage | - | Idealized Impact, Football | Ang. Acc. | GHBMC | (7) |
| BAM | 13 | Football, Naval Sled tests | Ang. Acc. | KTH | (8) |
| lin_acc_CG_max | - | - | Lin. Acc. | - | - |
| ang_vel_max | - | - | Ang. Vel. | - | - |
| ang_acc_max | - | - | Ang. Acc. | - | - |

*The formula for RVCI is based on the angular acceleration, but the integral of angular acceleration is the change of angular velocity.

### 2.3 Evaluation of brain strain predictability based on BIC

To evaluate the accuracy of estimating MPS95/MPSCC95/CSDM based on each of the 18 BIC, univariate ordinary least squares (OLS) regression models were built with each BIC as the predictor and the MPS95/MPSCC95/CSDM as the outcome, respectively. OLS regression was used as it indicates the most direct predictive power of BIC in terms of explained variance (coefficient of determination), and it has been the common practice used by the BIC developers in evaluating the accuracy of BIC

(6,8,21,26,33,40,41,42). Qualitatively, the residuals also generally meet the linear regression assumptions of relative constant residuals and normal distribution.

The univariate regression coefficients were initially analyzed to investigate the relationship between each BIC and the brain strain. For OLS, the two key components are the curve's slope and the intercept. Focus was placed on the slope because the slope determines the variance in the outcome that can be explained by the predictor. The intercept, which is in the same unit as the outcome, is used to balance the mean of the outcome in the regression to center the outcome data. To ensure robust results, the datasets were bootstrapped 1000 times (47). The mean and 95% confidence intervals of the regression slopes were recorded. The study for each of the 18 BIC was done independently with data standardization (to a mean of zero and a standard deviation of 1 for each BIC) before building the regression model.

Next, to further evaluate the cross-usage of BIC in different types of head impacts, the influence of different datasets on the accuracy of using BIC to estimate the MPS95/MPSCC95/CSDM was evaluated. The accuracy of brain strain estimation is recognized as higher if the coefficient of determination ($R^2$) of the OLS regression is higher (equivalent to that the Pearson coefficient of correlation (r) between the predictor and the outcome is higher in this single-predictor regression model).

## 2.4 Statistic analysis and regression tasks

In the analysis of the regression, the underlying relationship between the BIC and the brain strain across datasets were analyzed. Two statistical tests were performed: 1) the one-way ANOVA tests to find whether the regression slopes were statistically significantly different across datasets; 2) the pairwise Wilcoxon rank-sum tests (48) with Bonferroni correction to find statistical significance in the regression slope difference between each pair of two datasets. T-tests were not used because the Shapiro Wilk tests (49) rejected the normal distribution hypothesis on some of the regression slope results, and the Wilcoxon rank-sum test does not rely on the normal assumption of data distribution of the bootstrapped regression slope. Bonferroni correction was done on the significance level for the multiple tests and we only regarded p-values smaller than 0.005 as statistically significant. What must be mentioned here is that this study compared each BIC across datasets independently. Therefore, there are 10 pairwise tests for Bonferroni correction in the study of each of the 18 BIC.

Additionally, because the different regression slopes indicate the varying relationship between the BIC and the brain strain, the cross-dataset brain strain estimation accuracy may be different from that in the single-dataset prediction. To analyze the generalizability of the BIC model in estimating brain strain, three different tasks were designed:

1) <u>Single-dataset regression</u>: to test the ability of the BIC to estimate brain strain in the same type of head impacts, on each dataset, 80% of the impacts were randomly selected as the training dataset and the remaining 20% impacts as the testing dataset. The remaining 20% impacts were assumed to represent the distribution of the entire dataset from which they were randomly sampled. 100 iterations of bootstrapping were performed to ensure the result robustness. OLS models were fit on the training dataset and evaluated on the testing dataset, with the mean $R^2$ recorded to assess the accuracy. Five models corresponding to five datasets were built and evaluated.
2) <u>Cross-dataset regression</u>: in real-world applications, users usually apply BIC with published parameters trained on the development dataset to estimate the brain injury risks of new unseen impacts (33). To test the ability of the BIC to estimate brain strain on a different type of head impacts, a cross-dataset prediction task was conducted by using one dataset as the training dataset and another dataset as the testing dataset. 100 iterations of bootstrapping resampling were performed on the training dataset and the mean $R^2$ were recorded. Twenty models were built and evaluated because each of the five datasets could be either the training dataset or the testing dataset.

3) <u>Leave-one-dataset-out regression</u>: this task was completed to seek for potential improvement by combining different datasets in training BIC when the brain injury risk of an unseen type of head impact which needs to be estimated, which was employed in the development of BAM (8). To do this, four datasets were combined to be the training dataset with the remaining dataset as the testing dataset. The same pattern of bootstrapping resampling was conducted. Five models were built and evaluated, because each of the five datasets was once regarded as the testing dataset.

Based on these three regression tasks, we aim to demonstrate whether the BIC modeled on one type of head impacts can be directly applied to estimate brain strain for another head impact type with similar accuracy.

## 3. Results
### 3.1 Analysis of regression slope

To test whether the relationship between the BIC and the brain strain is the same across datasets with different head impact types, the slopes in the OLS regression models were analyzed with the outcomes (MPS95/MPSCC95/CSDM) on the predictor (each of the BIC, standardized). Firstly, in all the 270 regression models, the predictor (one of the 18 BIC) was statistically significantly associated with the response (MPS95/MPSCC95/CSDM) ($p<0.01$, t-test). Then, one-way ANOVA tests show statistical significance of the different slopes across datasets on each of the three outcome variables ($p<0.001$). **Fig. 3** and **Supplementary Fig. 1** show the regression slopes on the different datasets and the pairwise comparison Wilcoxon rank-sum test p-values with Bonferroni correction. The results show that 524 out of the 540 pairwise comparisons manifest statistical significance at the p-value threshold of 0.005 (Bonferroni correction), and 522 out of the 540 pairwise comparisons manifest statistical significance at the p-value threshold of 0.001. Although the regression slopes are all positive values, which means that larger values in BIC indicate higher brain strain, the statistically significant difference of regression slopes indicates that the exact relationship between the BIC and the brain strain varies with different types of head impacts.

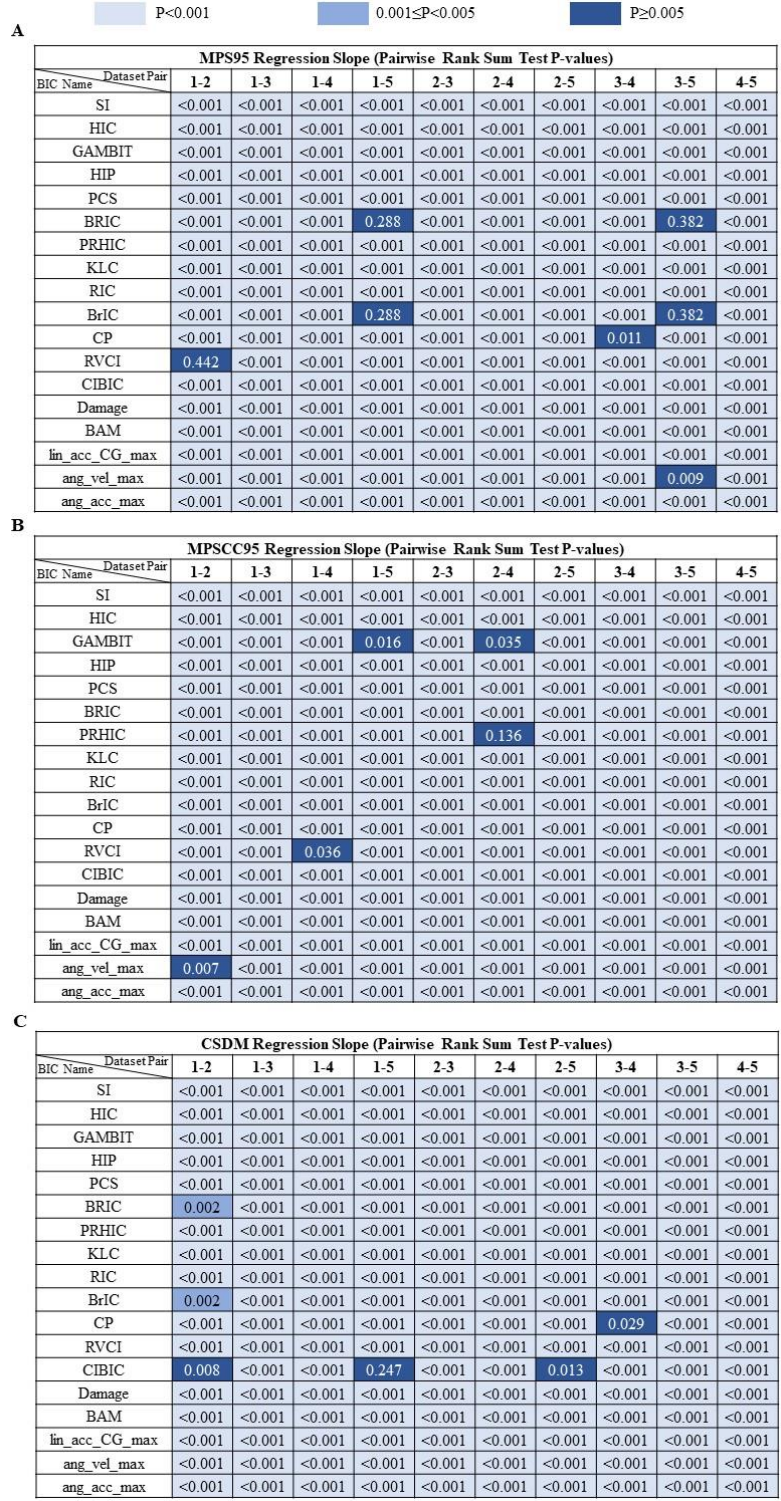

**Figure 3** The p-values of pairwise comparisons of regression slopes on different datasets of head impacts with 1000 iterations of bootstrapping resampling based on Wilcoxon rank-sum tests with Bonferroni correction. Dataset 1-5: lab impacts, college football, MMA, NHTSA, NASCAR. (A) The p-values of pairwise comparisons in the regression of MPS95. (B) The p-values of pairwise comparisons in the regression of MPSCC95. (C) The p-values of pairwise comparisons in the regression of CSDM.

### 3.2 Single-dataset regression (task 1) and cross-dataset regression (task 2)

As the analysis of regression slopes indicates the varying relationship between the BIC and the brain strain on different datasets, further testing was performed of the accuracy and generalizability of MPS95/MPSCC95/CSDM estimation on the single-dataset or across datasets. The diagonal elements in each plot of **Figs. 4-6** show the coefficients of determination ($R^2$) in single-dataset regression tasks, in which, the testing dataset comprised 20% data of the entire dataset while the training set comprised 80% data of the dataset to fit the OLS regression model. The off-diagonal elements in each plot of **Figs. 4-6** show the coefficients of determination ($R^2$) in cross-dataset regression in which one dataset was used as the training dataset and another dataset as the testing dataset for regression evaluation. The root mean squared error (RMSE) results are shown in **Supplementary Figs. 2-4**. The diagonal elements in each plot generally show the higher $R^2$ in single-dataset regression tasks than the off-diagonal elements which are the cross-dataset regression. There is also a high variance in the performance of different BIC.

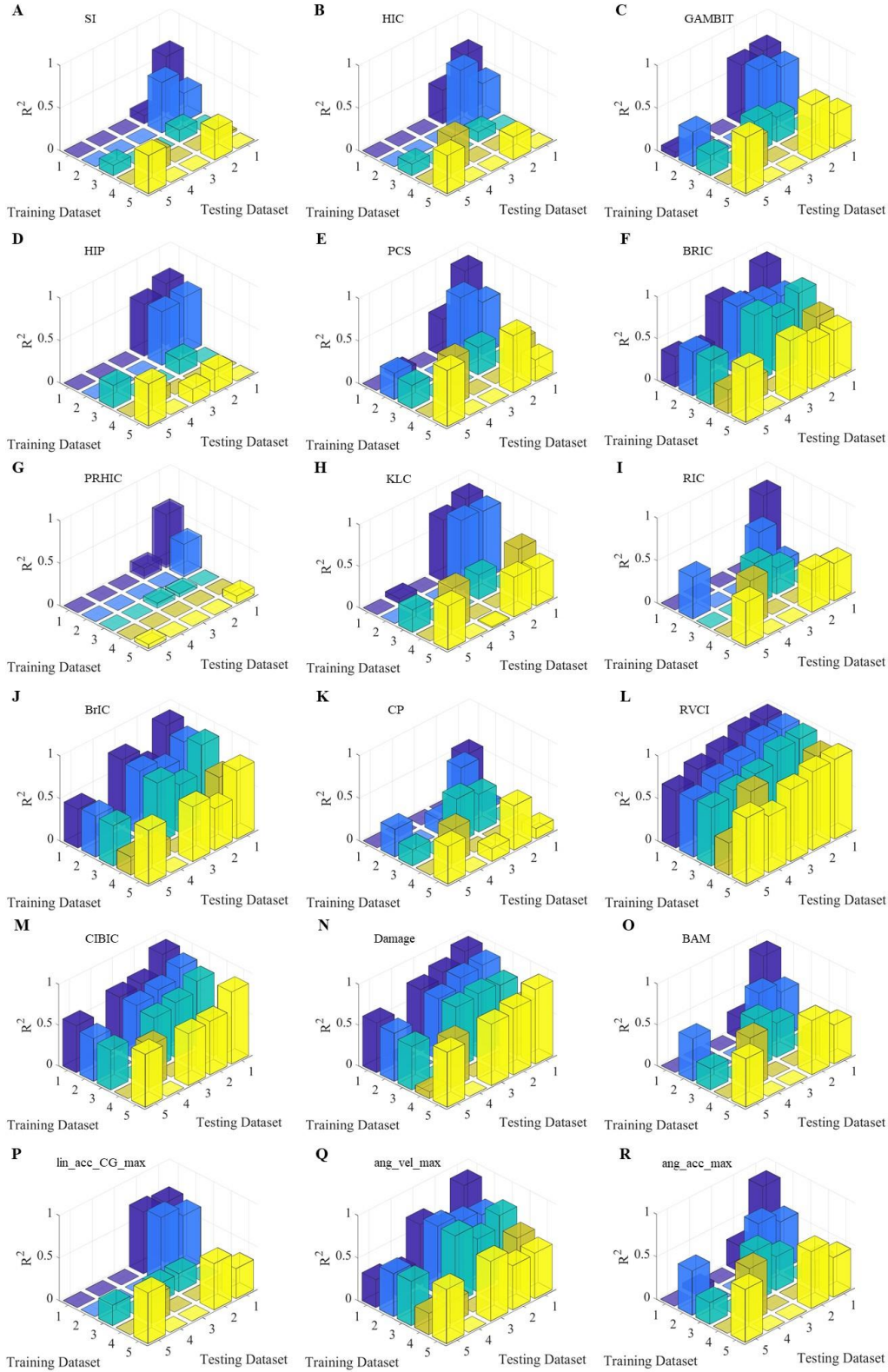

**Figure 4** The mean $R^2$ in the single-dataset regression and cross-dataset regression of MPS95 based on 18 BIC with 100 iterations of bootstrapping resampling. The color indicates the same training dataset. Dataset 1: lab impacts; Dataset 2: college football; Dataset 3: MMA; Dataset 4: NHTSA; Dataset5: NASCAR.

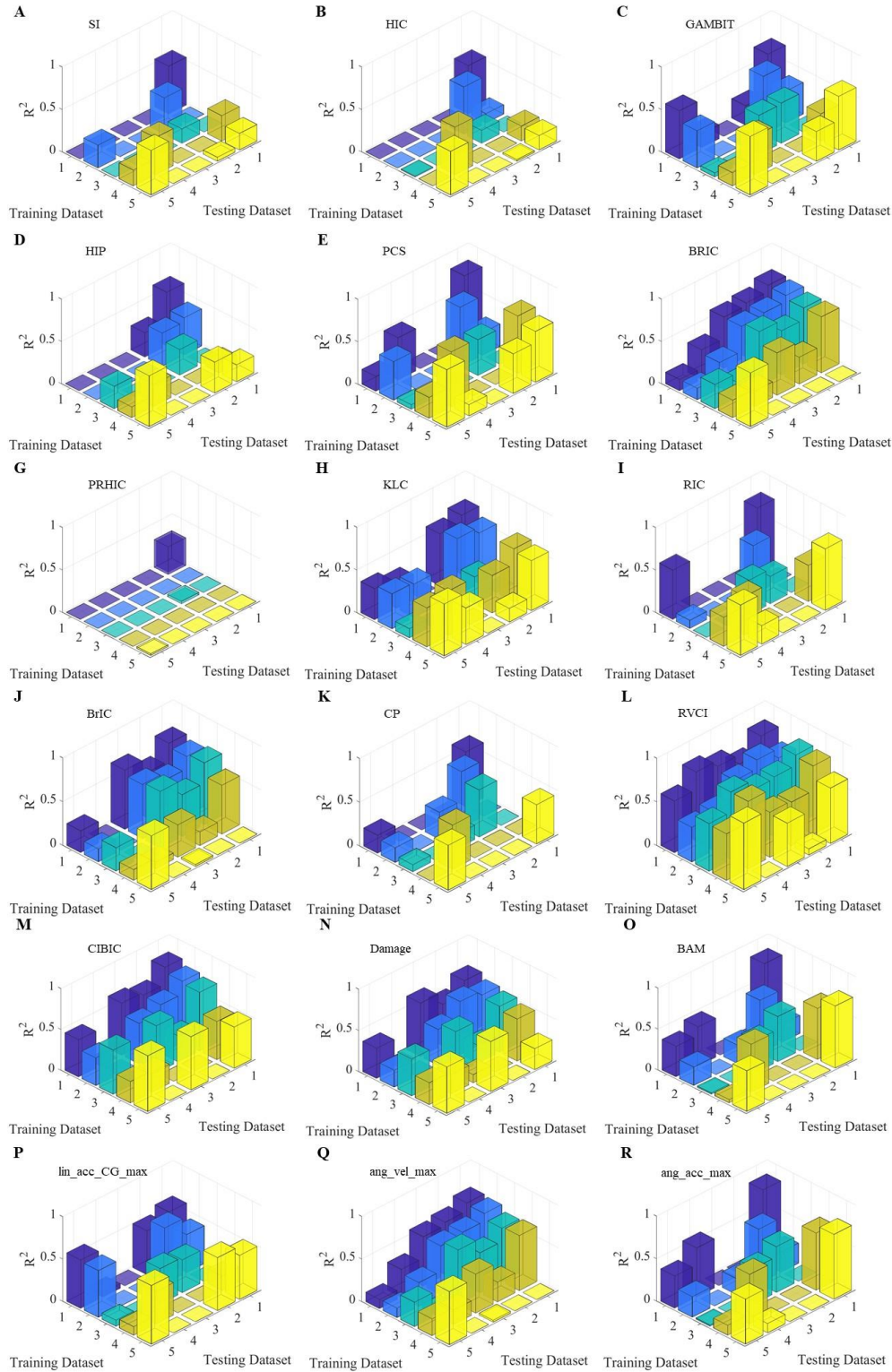

**Figure 5** The mean $R^2$ in the single-dataset regression and cross-dataset regression of MPSCC95 based on 18 BIC with 100 iterations of bootstrapping resampling. The color indicates the same training dataset. Dataset 1: lab impacts; Dataset 2: college football; Dataset 3: MMA; Dataset 4: NHTSA; Dataset5: NASCAR.

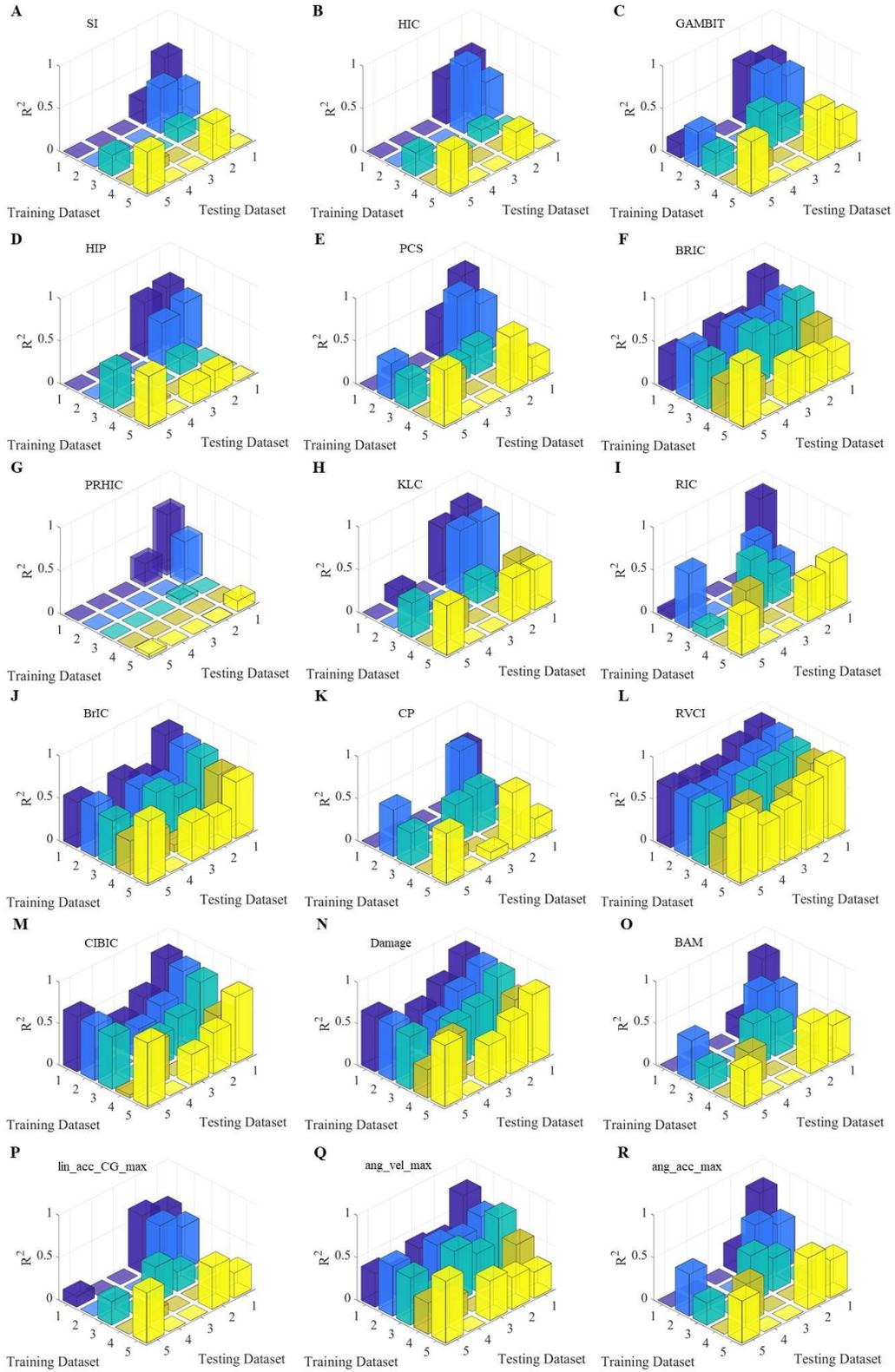

**Figure 6** The mean $R^2$ in the single-dataset regression and cross-dataset regression of CSDM based on 18 BIC with 100 iterations of bootstrapping resampling. The color indicates the same training dataset. Dataset 1: lab impacts; Dataset 2: college football; Dataset 3: MMA; Dataset 4: NHTSA; Dataset5: NASCAR.

### 3.3 Leave-one-dataset-out regression (task 3)

To seek for potential improvement by combining different datasets in training BIC when the brain injury risk of an unseen type of head impact which needs to be estimated, the leave-one-dataset-out regression was performed by leaving one dataset completely out as the testing datasets and the combination of remaining datasets as the training dataset. As is shown in **Fig. 7**, the mean $R^2$ was used as the accuracy metric, the accuracy of single-dataset regression tasks generally outperforms the accuracy of the leave-one-dataset-out regression tasks on the same dataset and with the same type of BIC. This result repeats on MPS95, MPSCC95 and CSDM regression, which shows that combining all the data not from the same type of head impacts in the testing dataset leads to lower accuracy in the brain strain regression.

## A

| Testing Set | 1-Lab Impacts | | 2-College Football | | 3-MMA | | 4-NHTSA | | 5-NASCAR | |
|---|---|---|---|---|---|---|---|---|---|---|
| BIC Name / Task | Single Dataset | Leave-one-out | Single Dataset | Leave-one-out | Single Dataset | Leave-one-out | Single Dataset | Leave-one-out | Single Dataset | Leave-one-out |
| SI | 0.658 | -0.196 | 0.591 | 0.063 | -2.815 | -10.096 | 0.220 | -0.946 | 0.448 | -0.202 |
| HIC | 0.671 | -0.187 | 0.734 | 0.025 | -3.730 | -8.566 | 0.424 | -1.737 | 0.447 | -0.558 |
| GAMBIT | 0.725 | 0.229 | 0.737 | 0.722 | 0.370 | -2.548 | 0.385 | -0.965 | 0.658 | 0.362 |
| HIP | 0.721 | -0.090 | 0.630 | 0.310 | -3.275 | -1.249 | 0.154 | -4.915 | 0.493 | -1.077 |
| PCS | 0.871 | 0.113 | 0.797 | 0.647 | -0.312 | -5.795 | 0.548 | -0.517 | 0.657 | 0.128 |
| BRIC | 0.906 | 0.820 | 0.739 | 0.606 | 0.806 | 0.793 | 0.270 | 0.327 | 0.650 | 0.422 |
| PRHIC | 0.624 | -0.240 | -0.769 | -0.341 | 0.060 | -169.414 | -1.776 | -2.922 | 0.049 | -0.098 |
| KLC | 0.862 | 0.387 | 0.840 | 0.617 | -2.907 | -3.087 | 0.499 | 0.276 | 0.526 | 0.110 |
| RIC | 0.814 | -0.124 | 0.622 | 0.240 | 0.496 | -21.961 | 0.526 | -1.116 | 0.512 | 0.338 |
| BrIC | 0.908 | 0.880 | 0.628 | 0.557 | 0.709 | 0.771 | 0.199 | -0.826 | 0.630 | 0.469 |
| CP | 0.554 | -0.005 | 0.644 | 0.224 | 0.518 | -1.620 | 0.398 | -1.809 | 0.438 | 0.004 |
| RVCI | 0.973 | 0.967 | 0.972 | 0.958 | 0.815 | 0.865 | 0.848 | 0.789 | 0.775 | 0.671 |
| CIBIC | 0.918 | 0.848 | 0.722 | 0.675 | 0.615 | 0.656 | 0.513 | -3.272 | 0.646 | 0.580 |
| Damage | 0.954 | 0.923 | 0.868 | 0.817 | 0.748 | 0.748 | 0.573 | -1.565 | 0.671 | 0.607 |
| BAM | 0.890 | 0.278 | 0.675 | 0.609 | 0.588 | -2.836 | 0.604 | -0.942 | 0.589 | 0.283 |
| lin_acc_CG_max | 0.703 | 0.232 | 0.766 | 0.664 | 0.209 | -2.163 | 0.314 | -1.082 | 0.596 | 0.267 |
| ang_vel_max | 0.903 | 0.809 | 0.713 | 0.581 | 0.788 | 0.764 | 0.256 | 0.300 | 0.635 | 0.394 |
| ang_acc_max | 0.876 | 0.267 | 0.676 | 0.628 | 0.591 | -2.965 | 0.626 | -0.742 | 0.619 | 0.301 |

## B

| Testing Set | 1-Lab Impacts | | 2-College Football | | 3-MMA | | 4-NHTSA | | 5-NASCAR | |
|---|---|---|---|---|---|---|---|---|---|---|
| BIC Name / Task | Single Dataset | Leave-one-out | Single Dataset | Leave-one-out | Single Dataset | Leave-one-out | Single Dataset | Leave-one-out | Single Dataset | Leave-one-out |
| SI | 0.562 | -0.105 | 0.427 | -0.673 | -2.270 | -14.653 | 0.435 | -0.142 | 0.522 | 0.338 |
| HIC | 0.568 | -0.091 | 0.555 | -0.649 | -2.881 | -12.854 | 0.558 | -0.544 | 0.511 | 0.289 |
| GAMBIT | 0.702 | 0.296 | 0.679 | 0.453 | 0.459 | -3.791 | 0.451 | 0.038 | 0.686 | 0.550 |
| HIP | 0.633 | 0.019 | 0.389 | 0.000 | -3.235 | -1.476 | 0.258 | -3.921 | 0.598 | 0.260 |
| PCS | 0.827 | 0.186 | 0.696 | 0.136 | -0.107 | -8.662 | 0.647 | 0.237 | 0.675 | 0.529 |
| BRIC | 0.715 | 0.627 | 0.628 | 0.593 | 0.644 | 0.669 | 0.183 | 0.348 | 0.633 | 0.163 |
| PRHIC | 0.325 | -0.190 | -0.776 | -1.325 | -0.184 | -132.657 | -1.023 | -1.337 | 0.018 | 0.004 |
| KLC | 0.692 | 0.368 | 0.657 | 0.370 | -2.533 | -4.174 | 0.521 | 0.431 | 0.606 | 0.494 |
| RIC | 0.781 | -0.061 | 0.576 | -0.726 | 0.406 | -30.663 | 0.453 | -0.272 | 0.591 | 0.337 |
| BrIC | 0.719 | 0.700 | 0.534 | 0.454 | 0.644 | 0.687 | 0.073 | -0.419 | 0.613 | 0.243 |
| CP | 0.628 | 0.122 | 0.640 | -0.839 | 0.161 | -2.519 | 0.505 | -0.299 | 0.512 | 0.455 |
| RVCI | 0.808 | 0.819 | 0.772 | 0.601 | 0.665 | 0.722 | 0.773 | 0.790 | 0.761 | 0.584 |
| CIBIC | 0.813 | 0.764 | 0.591 | 0.535 | 0.572 | 0.657 | 0.234 | -2.319 | 0.681 | 0.445 |
| Damage | 0.656 | 0.656 | 0.641 | 0.445 | 0.513 | 0.646 | 0.234 | -0.670 | 0.562 | 0.351 |
| BAM | 0.850 | 0.324 | 0.641 | -0.104 | 0.388 | -4.284 | 0.581 | 0.044 | 0.495 | 0.442 |
| lin_acc_CG_max | 0.608 | 0.276 | 0.656 | 0.515 | 0.416 | -2.950 | 0.381 | -0.068 | 0.691 | 0.554 |
| ang_vel_max | 0.709 | 0.613 | 0.603 | 0.566 | 0.634 | 0.652 | 0.169 | 0.325 | 0.622 | 0.142 |
| ang_acc_max | 0.861 | 0.323 | 0.637 | -0.046 | 0.418 | -4.515 | 0.582 | 0.161 | 0.529 | 0.452 |

## C

| Testing Set | 1-Lab Impacts | | 2-College Football | | 3-MMA | | 4-NHTSA | | 5-NASCAR | |
|---|---|---|---|---|---|---|---|---|---|---|
| BIC Name / Task | Single Dataset | Leave-one-out | Single Dataset | Leave-one-out | Single Dataset | Leave-one-out | Single Dataset | Leave-one-out | Single Dataset | Leave-one-out |
| SI | 0.642 | -0.108 | 0.528 | 0.149 | -1.842 | -14.760 | 0.121 | -0.835 | 0.494 | 0.011 |
| HIC | 0.644 | -0.101 | 0.787 | 0.107 | -2.587 | -12.228 | 0.215 | -1.456 | 0.506 | -0.279 |
| GAMBIT | 0.619 | 0.212 | 0.699 | 0.687 | 0.487 | -3.315 | 0.192 | -0.949 | 0.615 | 0.394 |
| HIP | 0.676 | -0.011 | 0.499 | 0.344 | -1.791 | -1.475 | 0.029 | -3.081 | 0.589 | -0.761 |
| PCS | 0.814 | 0.159 | 0.810 | 0.654 | 0.249 | -8.201 | 0.362 | -0.548 | 0.657 | 0.268 |
| BRIC | 0.787 | 0.702 | 0.537 | 0.467 | 0.626 | 0.582 | 0.273 | 0.301 | 0.734 | 0.500 |
| PRHIC | 0.704 | -0.150 | -1.737 | -0.272 | -0.005 | -280.909 | -0.993 | -2.253 | 0.037 | -0.082 |
| KLC | 0.770 | 0.382 | 0.745 | 0.580 | -1.464 | -4.309 | 0.328 | 0.111 | 0.581 | 0.403 |
| RIC | 0.885 | -0.023 | 0.630 | 0.370 | 0.642 | -36.709 | 0.514 | -0.872 | 0.468 | 0.441 |
| BrIC | 0.793 | 0.769 | 0.443 | 0.428 | 0.595 | 0.575 | 0.333 | 0.011 | 0.734 | 0.560 |
| CP | 0.652 | 0.127 | 0.851 | 0.400 | 0.452 | -2.565 | 0.133 | -1.647 | 0.590 | 0.252 |
| RVCI | 0.897 | 0.884 | 0.829 | 0.817 | 0.745 | 0.670 | 0.701 | 0.674 | 0.770 | 0.706 |
| CIBIC | 0.828 | 0.753 | 0.504 | 0.499 | 0.359 | 0.378 | 0.518 | -1.121 | 0.769 | 0.663 |
| Damage | 0.865 | 0.829 | 0.678 | 0.634 | 0.519 | 0.492 | 0.588 | -0.268 | 0.740 | 0.655 |
| BAM | 0.824 | 0.285 | 0.675 | 0.578 | 0.521 | -4.178 | 0.411 | -0.852 | 0.438 | 0.259 |
| lin_acc_CG_max | 0.589 | 0.208 | 0.665 | 0.610 | 0.416 | -2.738 | 0.122 | -1.044 | 0.590 | 0.343 |
| ang_vel_max | 0.783 | 0.690 | 0.509 | 0.442 | 0.604 | 0.550 | 0.262 | 0.286 | 0.727 | 0.476 |
| ang_acc_max | 0.811 | 0.278 | 0.671 | 0.596 | 0.551 | -4.346 | 0.455 | -0.694 | 0.495 | 0.290 |

Color legend: $R^2 \geq 0.80$; $0.60 \leq R^2 < 0.80$; $0.40 \leq R^2 < 0.60$; $0.20 \leq R^2 < 0.40$; $0.00 \leq R^2 < 0.20$; $R^2 < 0$

**Figure 7** Comparison of mean $R^2$ in the single-dataset regression task and the leave-one-dataset-out regression task on each dataset and each BIC with 100 iterations of bootstrapping resampling. (A) The mean $R^2$ in the regression of MPS95. (B) The mean $R^2$ in the regression of MPSCC95. (C) The mean $R^2$ in the regression of CSDM.

## 4. Discussion

Although different BIC have been developed based on different types of head impacts, the accuracy of using the BIC to estimate brain injury risks across different types of head impacts has previously not been investigated. This study investigated the regression slopes in the OLS regression models of MPS95/MPSCCC95/CSDM on 18 BIC and the generalizability of the regression models to estimate the brain strain if the models are fitted on a different dataset, with five datasets of different types of head impacts (lab impacts, college football impacts, MMA impacts, NHTSA impacts, NASCAR impacts). This study found that the regression slopes in most of the OLS regression models showed statistically significantly differences across datasets in **Fig. 3** and **Supplementary Fig. 1**. This indicates that the underlying relationship between BIC and brain strain varies with different types of head impacts. For example, if two impacts (one from football and one from car crash) reached the same value of BIC, the brain injury risks for these two impacts can still be very different. A potential explanation is that the characteristics of the kinematics used by BIC to estimate brain strain vary among different types of head impacts. Furthermore, the difference in the relationship between BIC and brain strain suggests that the risks of brain injury vary among different types of head impacts.

The influence of cross-using BIC across head impact types on the accuracy of the brain strain estimation was shown in **Figs. 4-6** and **Supplementary Figs. 2-4**. The bars were generally higher on the diagonal than off diagonal, indicating that the single-dataset regression tasks generally showed higher mean $R^2$ than those in the cross-dataset regression tasks. In other words, the models were more accurate in the regression of MPS95/MPSCC95/CSDM if it was fit on the same type of head impacts. It should be noted that exceptions did exist when the off-diagonal regression accuracy was higher, and the potential reasons may be the insignificant slope difference in the regression of the two datasets and the randomness in a small dataset. This finding suggests that the accuracy of the BIC should be evaluated according to the different types of head impact respectively. In previous studies related to BIC, the evaluation of the accuracy in brain strain regression has been mainly done on the same dataset. Although different types of head impacts were combined to develop BIC in previous studies (**Table.1**), it was only until recently that researchers started to investigate the declining performance of the brain strain regression models when applied to a different type of head impacts. Zhan et al (20) developed a deep learning head model for predicting brain strain of the entire brain and showed that the model accuracy deteriorates when the model was trained on a dataset mainly comprising simulated head impacts and applied to MMA head impacts. According to the finding in this study, concerns may arise if the BIC is used to estimate the brain strain if the BIC regression model is fit on a different type of head impacts. Therefore, the BIC, the reduced-order head models and deep learning head models developed by researchers to predict brain injury risks need to be validated across different types of head impacts if they are ever intended to be used that way.

Additionally, although this study found statistically significant different regression slopes across datasets in **Fig. 3**, the findings also showed that there might be similarities between certain pairs of head impact types. In terms of the regression slopes in **Fig. 3**, on BrIC and BRIC, statistical significance was not found in the pairwise comparisons between lab impacts and NASCAR impacts and between MMA impacts and NASCAR impacts, in the regression of MPS95. On CP, no statistical significance was found in the pairwise comparison between MMA impacts and NHTSA impacts in the regression of MPS95 and CSDM. On RVCI, no statistical significance was found in the regression of MPS95 in the pairwise comparisons between lab impacts and college football impacts; no statistical significance was found in the regression of MPSCC95 in the pairwise comparison between lab impacts and NHTSA impacts. On CIBIC, no statistical significance was found in the pairwise comparisons between lab impacts and college football impacts, between lab impacts and NASCAR impacts, and between college football impacts and NASCAR impacts, in the regression of CSDM. On GAMBIT, in the regression of MPSCC95, no statistical significance was found between lab impacts and NASCAR impacts and between college football impacts and NHTSA impacts. On PRHIC, no statistical significance was found between college football impacts and NHTSA impacts in MPSCC95 regression. These similarities among different types of head impacts can also be indicated in the cross-dataset regression results in which there were several high bars off the diagonal. For example, high mean $R^2$ was shown in the pairs (training dataset listed first in each pair): NASCAR impacts-lab impacts, NASCAR impacts-college football impacts, college football impacts-NASCAR impacts, college football impacts-lab impacts, and MMA impacts-college football impacts. The possible explanations may be: 1) the lab impacts were generally simulated for the research of head impacts in football matches, which may explain the similarity between lab impacts and college football impacts; 2) the MMA head impacts and NHTSA car crash dummy head impacts were measured without protective helmets, while the college football head impacts and the NASCAR head impacts were measured with protective helmets that protect the players/drivers from concussion. The helmet can effectively change the impact response of the head (50), which may explain the similarity between MMA impacts and NHTSA impacts, as well as among NASCAR impacts, college football impacts and lab impacts.

Comparing the performances of BIC in single-dataset task in **Figs. 4-6**, most of BIC definitions can accurately estimate brain strain, and the BIC that included angular velocity as input (BRIC, BrIC, RVCI, ang_vel_max) as well as the BIC developed by solving the mass-lumped systems (Damage, CIBIC. BAM) generally showed higher accuracy (the high diagonal bars), which agrees with the results published in (7,15,51). Furthermore, these two types of BIC also provided the generally accurate regression in the cross-dataset task (the high off diagonal bars) except for BAM. The different performances of BIC in cross-dataset tasks derive from the fact that each BIC was designed to capture certain characteristics of the head kinematics to estimate the brain strain, and these characteristics varied among different datasets. Although the mechanism underlying how the head kinematics caused the brain strain was not clear, a potential factor might influence the performance in the cross-dataset task was the impulse duration. As observed in (6) and explained in (16), the MPS95 depends on the angular velocity peak when the impulse duration was short, and on the angular acceleration peak when the impulse duration was longer. The impulse duration was only defined for idealized impact instead of the on-field impact. It is possible that the difference in impulse duration contributed to the difference in BIC performances. Aside from the impulse duration, the multiple peaks (**Fig.1A**) and the different profiles (51–53) were also potential factors that influenced the cross-dataset performance.

The lower accuracy in leave-one-dataset-out regression in **Fig.7** suggested that combining several different types of head impacts as the training dataset may fail to estimate the brain strain in the unknown type of head impacts. As shown in **Table.1**, most of the BIC were developed based on the football and traffic accident datasets. Therefore, researchers and medical professionals should carefully interpret the results of applying BIC to other type of head impacts. For the type of head impact in which the head kinematics characteristic is unknown, sufficient impact cases should be collected and analyzed by an FE model before using the BIC. Furthermore, deep learning head models were recently developed based on football dataset (54) and football, MMA, and lab impact datasets (20), and these models have achieved promising accuracy. Although different types of dataset were used in (20), the ability of the model to predict the brain strain in an unknown type of head impact was not examined.

Furthermore, multiple negative mean $R^2$ occurred in this study in **Fig. 4-7**, mainly in the cross-dataset regression tasks and leave-one-dataset-out regression tasks but also in the single-dataset regression tasks. The negative mean $R^2$ in the results may be explained by the significantly different distributions of the training dataset and the test dataset. In the cross-datasets and leave-one-dataset-out tasks, the negative values indicate that the distribution of the other datasets may be insufficient to represent the distribution of the selected testing dataset so that the regression models showed high bias. In the single-dataset tasks, the negative values may be caused by the fact that the random sampled 20% testing data may not be well covered by some of the bootstrapping resampled training dataset distributions. For instance, the MMA dataset had two peaks with significantly different kernel densities, which indicates that there are much fewer data points in the minor peak region (as is shown in the red box in **Fig. 2**). These may be caused by several exceptionally fierce fights with severe head impacts and the collection of more data may address this issue. The data points located in the minor peak of the MPS95/MPSCC95 distributions of the MMA dataset in the testing dataset might not be well covered by the distribution of the bootstrapping resampled training dataset because the bootstrapping resampled training datasets theoretically covers 63.2% samples in the training data (55). The NHTSA dataset, which only contains 48 head impacts, may also suffer from the distributional mismatch between the testing dataset and the training dataset.

In addition, the decrease in the regression accuracy from single-dataset tasks to cross-dataset tasks and leave-one-dataset-out tasks also indicates the issue of overfitting of the BIC models, even with simple OLS regression. Because data fitting is widely used to assess BIC accuracy (e.g., fitting a regression model of 95% MPS on HIC) by evaluating the goodness of fit, however, to fit the model and to evaluate the goodness of fit on the same dataset may cause overfitting. The overfitting can lead to over-optimistic estimate of the accuracy of the BIC. Therefore, instead of fitting the BIC and evaluating the goodness of fit on the same dataset, the accuracy of the BIC across different head impact datasets also should to be validated.

There are a few important limitations of this study to note: firstly, most of the BIC definitions were developed with parameters by optimizing the brain injury risk estimation in certain datasets (**Table.1**). In this study, the BIC values were used directly with the parameters reported in the papers. It is possible that recalculating the parameters according to each dataset may probably generate higher accuracy than what was reported in this paper. Therefore, it is suggested that each BIC should be further improved for different types of head impacts in the future.

Additionally, we used KTH model (32) as the finite element head model (FEHM), which is relatively simplified compared to the current state-of-the-art FEHM (56,57). The KTH model has tangential sliding without separation in the normal direction between subdural cerebral spinal fluid and brain. In addition, there are no gyri or sulci, which have been confirmed to have a significant influence on the FEHM behavior and its intracranial motion. The KTH model also lacks the interface between white and grey matter. In the future, new FEHM developed in recent years can be applied and tested.

Also, it was assumed that the testing dataset represented the head kinematics in the type of head impacts. This assumption should be examined particularly when the dataset distribution is skewed with multiple peaks or heavy tails. In the future, as more head impact data becomes available, this assumption will be more robust. Finally, the OLS regression model was the only method used to predict MPS95/MPSCC95/CSDM from the BIC because it directly shows the ability of brain strain estimation based on BIC; however, the underlying relationship between BIC and the brain strain may also be non-linear (25). The simple linear regression may have certain biases. Neural-network-based classifiers may perform better across the cohort of comparisons listed here but may even more drastically overfit the data. Lastly, we regarded the ability to predict the brain strain as the ability to evaluate brain injury risk. More neurological studies need to be done to validate the brain strain as an accurate supplant of brain injury risk.

## 5. Conclusion

In this study, we investigated the varying relationship between the BIC and the brain strain and the different accuracy of brain strain estimation over different datasets of head impacts. The slope values of linear regression from each of the 18 BIC to brain strain (MPS95/MPSCC95/CSDM) manifest statistically significant differences across datasets. Furthermore, the accuracy of the estimation of MPS95/MPSCC95/CSDM on each BIC may be lower if the regression models are fitted on different training datasets and then used to estimate the brain strain on another testing dataset. The findings indicate researchers and medical professionals who use the BIC to monitor the severity of head impact and to estimate the brain strain to take extra caution to interpret the BIC results in different types of head impact. Based on the findings, it may also be advantageous to develop type-specific BIC brain strain estimation models on the same type of head impacts as the head impacts which need to be evaluated.

**Acknowledgements**

The authors want to thank Dr. Kaveh Laksari, Dr. Lee F. Gabler and Dr. Toshiyuki Yanaoka for answering questions and providing help. This work is also supported by Stanford Department of Bioengineering.

**Funding**

This research was supported by the Pac-12 Conference's Student-Athlete Health and Well-Being Initiative, the National Institutes of Health (R24NS098518), Taube Stanford Children's Concussion Initiative.

**Supplementary Materials**
**Data and Code**
The data and codes used in this study are reported at: https://github.com/xzhan96-stf/BIC.

**Supplementary Figures**

### A

| BIC Name | Dataset1 (mean, 95%CI) | Dataset2 (mean, 95%CI) | Dataset3 (mean, 95%CI) | Dataset4 (mean, 95%CI) | Dataset5 (mean, 95%CI) |
|---|---|---|---|---|---|
| SI | 1.861e-01 [1.800e-01,1.926e-01] | 1.099e-01 [8.853e-02,1.781e-01] | 2.001e-02 [8.252e-03,4.790e-02] | 5.008e-02 [3.377e-02,7.182e-02] | 3.701e-02 [2.775e-02,5.148e-02] |
| HIC | 1.870e-01 [1.806e-01,1.935e-01] | 1.322e-01 [1.018e-01,2.118e-01] | 2.065e-02 [7.800e-03,4.868e-02] | 6.424e-02 [4.562e-02,9.152e-02] | 3.234e-02 [2.374e-02,4.603e-02] |
| GAMBIT | 9.081e-02 [8.838e-02,9.320e-02] | 7.483e-02 [6.493e-02,8.456e-02] | 3.100e-02 [2.401e-02,3.857e-02] | 4.161e-02 [2.938e-02,5.832e-02] | 6.306e-02 [5.334e-02,7.249e-02] |
| HIP | 1.237e-01 [1.204e-01,1.273e-01] | 1.292e-01 [1.080e-01,1.567e-01] | 3.138e-02 [1.384e-02,6.370e-02] | 1.420e-02 [6.192e-03,2.042e-02] | 3.113e-02 [2.365e-02,3.984e-02] |
| PCS | 1.214e-01 [1.191e-01,1.238e-01] | 7.600e-02 [6.728e-02,8.599e-02] | 2.672e-02 [1.653e-02,3.970e-02] | 6.587e-02 [5.256e-02,8.698e-02] | 5.243e-02 [4.153e-02,6.617e-02] |
| BRIC | 8.906e-02 [8.775e-02,9.040e-02] | 1.092e-01 [8.983e-02,1.273e-01] | 9.937e-02 [8.867e-02,1.081e-01] | 5.808e-02 [3.346e-02,8.143e-02] | 1.039e-01 [7.624e-02,1.329e-01] |
| PRHIC | 3.415e-01 [3.210e-01,3.624e-01] | 3.192e-01 [1.889e-01,7.225e-01] | 4.651e-02 [1.294e-02,3.160e-01] | 2.204e-01 [8.935e-02,5.019e-01] | 1.374e-01 [4.882e-03,4.696e-01] |
| KLC | 1.107e-01 [1.085e-01,1.129e-01] | 1.146e-01 [1.009e-01,1.370e-01] | 3.154e-02 [1.346e-02,6.687e-02] | 6.232e-02 [4.253e-02,8.179e-02] | 4.581e-02 [3.520e-02,6.137e-02] |
| RIC | 1.813e-01 [1.736e-01,1.890e-01] | 6.893e-02 [5.823e-02,8.926e-02] | 2.079e-02 [1.681e-02,2.772e-02] | 7.799e-02 [6.301e-02,1.138e-01] | 1.000e-01 [8.166e-02,1.265e-01] |
| BrIC | 8.930e-02 [8.799e-02,9.063e-02] | 1.003e-01 [8.010e-02,1.210e-01] | 8.827e-02 [7.569e-02,1.001e-01] | 4.681e-02 [2.701e-02,7.033e-02] | 8.867e-02 [5.813e-02,1.181e-01] |
| CP | 8.833e-02 [8.245e-02,9.467e-02] | 4.690e-02 [3.697e-02,5.660e-02] | 2.945e-02 [2.211e-02,3.774e-02] | 2.990e-02 [1.982e-02,3.976e-02] | 3.590e-02 [2.702e-02,4.367e-02] |
| RVCI | 9.049e-02 [8.981e-02,9.120e-02] | 9.067e-02 [8.596e-02,9.589e-02] | 8.822e-02 [7.805e-02,9.821e-02] | 7.095e-02 [6.196e-02,8.272e-02] | 8.553e-02 [6.849e-02,9.682e-02] |
| CIBIC | 9.061e-02 [8.902e-02,9.215e-02] | 1.053e-01 [8.806e-02,1.214e-01] | 1.089e-01 [9.121e-02,1.277e-01] | 3.042e-02 [2.020e-02,3.824e-02] | 8.244e-02 [5.963e-02,9.986e-02] |
| Damage | 9.037e-02 [8.930e-02,9.145e-02] | 1.068e-01 [9.507e-02,1.169e-01] | 1.157e-01 [1.013e-01,1.296e-01] | 4.054e-02 [3.262e-02,4.938e-02] | 9.274e-02 [6.733e-02,1.138e-01] |
| BAM | 1.014e-01 [9.972e-02,1.031e-01] | 5.904e-02 [4.937e-02,6.805e-02] | 3.298e-02 [2.663e-02,3.950e-02] | 6.983e-02 [5.680e-02,8.583e-02] | 6.501e-02 [5.604e-02,7.486e-02] |
| lin_acc_CG_max | 8.674e-02 [8.417e-02,8.909e-02] | 9.326e-02 [8.166e-02,1.060e-01] | 2.910e-02 [2.103e-02,3.808e-02] | 3.488e-02 [2.329e-02,4.708e-02] | 5.583e-02 [4.559e-02,6.572e-02] |
| ang_vel_max | 8.913e-02 [8.780e-02,9.055e-02] | 1.086e-01 [8.843e-02,1.275e-01] | 1.024e-01 [9.053e-02,1.121e-01] | 5.676e-02 [3.180e-02,8.045e-02] | 1.042e-01 [7.595e-02,1.345e-01] |
| ang_acc_max | 1.012e-01 [9.934e-02,1.031e-01] | 6.026e-02 [5.055e-02,6.936e-02] | 3.276e-02 [2.630e-02,3.953e-02] | 7.036e-02 [6.010e-02,8.312e-02] | 6.445e-02 [5.432e-02,7.468e-02] |

### B

| BIC Name | Dataset1 (mean, 95%CI) | Dataset2 (mean, 95%CI) | Dataset3 (mean, 95%CI) | Dataset4 (mean, 95%CI) | Dataset5 (mean, 95%CI) |
|---|---|---|---|---|---|
| SI | 2.305e-01 [2.200e-01,2.408e-01] | 1.005e-01 [7.162e-02,1.804e-01] | 2.018e-02 [1.152e-02,5.414e-02] | 9.338e-02 [6.225e-02,1.500e-01] | 7.612e-02 [5.533e-02,1.074e-01] |
| HIC | 2.302e-01 [2.197e-01,2.404e-01] | 1.238e-01 [8.432e-02,2.255e-01] | 2.097e-02 [1.112e-02,5.515e-02] | 1.127e-01 [7.485e-02,1.634e-01] | 6.819e-02 [4.847e-02,9.976e-02] |
| GAMBIT | 1.201e-01 [1.165e-01,1.237e-01] | 7.167e-02 [5.916e-02,8.618e-02] | 3.368e-02 [2.779e-02,4.156e-02] | 7.571e-02 [5.360e-02,1.135e-01] | 1.211e-01 [9.987e-02,1.436e-01] |
| HIP | 1.556e-01 [1.497e-01,1.613e-01] | 1.239e-01 [9.873e-02,1.656e-01] | 3.859e-02 [2.051e-02,6.983e-02] | 2.586e-02 [1.332e-02,3.985e-02] | 6.244e-02 [4.657e-02,8.121e-02] |
| PCS | 1.608e-01 [1.569e-01,1.646e-01] | 7.112e-02 [5.793e-02,8.698e-02] | 2.786e-02 [2.026e-02,4.438e-02] | 1.134e-01 [8.004e-02,1.585e-01] | 9.866e-02 [7.593e-02,1.265e-01] |
| BRIC | 1.079e-01 [1.047e-01,1.110e-01] | 1.070e-01 [8.640e-02,1.278e-01] | 1.112e-01 [9.620e-02,1.304e-01] | 9.007e-02 [5.862e-02,1.198e-01] | 1.790e-01 [1.219e-01,2.473e-01] |
| PRHIC | 3.448e-01 [3.163e-01,3.775e-01] | 3.011e-01 [1.682e-01,8.265e-01] | 4.974e-02 [7.358e-03,4.561e-01] | 2.940e-01 [6.879e-02,6.886e-01] | 2.301e-01 [8.339e-03,8.872e-01] |
| KLC | 1.345e-01 [1.300e-01,1.388e-01] | 1.087e-01 [8.544e-02,1.443e-01] | 3.327e-02 [1.838e-02,7.804e-02] | 1.055e-01 [7.619e-02,1.476e-01] | 9.045e-02 [6.848e-02,1.220e-01] |
| RIC | 2.427e-01 [2.343e-01,2.521e-01] | 6.500e-02 [4.987e-02,9.341e-02] | 1.983e-02 [1.223e-02,3.244e-02] | 1.170e-01 [9.253e-02,1.936e-01] | 1.954e-01 [1.567e-01,2.455e-01] |
| BrIC | 1.082e-01 [1.052e-01,1.111e-01] | 9.791e-02 [7.638e-02,1.201e-01] | 1.006e-01 [9.049e-02,1.108e-01] | 7.208e-02 [4.529e-02,1.039e-01] | 1.539e-01 [9.952e-02,2.164e-01] |
| CP | 1.245e-01 [1.160e-01,1.331e-01] | 4.639e-02 [3.680e-02,5.837e-02] | 3.260e-02 [2.463e-02,4.105e-02] | 5.258e-02 [4.139e-02,6.708e-02] | 7.416e-02 [5.498e-02,9.083e-02] |
| RVCI | 1.133e-01 [1.102e-01,1.161e-01] | 8.229e-02 [7.176e-02,9.779e-02] | 9.915e-02 [8.695e-02,1.143e-01] | 1.144e-01 [8.766e-02,1.465e-01] | 1.547e-01 [1.211e-01,1.816e-01] |
| CIBIC | 1.158e-01 [1.128e-01,1.186e-01] | 1.016e-01 [8.269e-02,1.205e-01] | 1.317e-01 [1.185e-01,1.454e-01] | 4.452e-02 [2.967e-02,6.284e-02] | 1.512e-01 [1.076e-01,1.932e-01] |
| Damage | 1.032e-01 [9.996e-02,1.063e-01] | 9.483e-02 [7.939e-02,1.122e-01] | 1.282e-01 [1.145e-01,1.443e-01] | 5.694e-02 [4.221e-02,7.860e-02] | 1.528e-01 [1.032e-01,2.048e-01] |
| BAM | 1.358e-01 [1.327e-01,1.387e-01] | 5.566e-02 [4.585e-02,6.648e-02] | 3.467e-02 [2.852e-02,4.222e-02] | 1.175e-01 [9.191e-02,1.534e-01] | 1.104e-01 [8.983e-02,1.330e-01] |
| lin_acc_CG_max | 1.080e-01 [1.039e-01,1.119e-01] | 8.975e-02 [7.357e-02,1.092e-01] | 3.214e-02 [2.681e-02,3.979e-02] | 6.518e-02 [4.564e-02,9.792e-02] | 1.137e-01 [9.215e-02,1.360e-01] |
| ang_vel_max | 1.077e-01 [1.044e-01,1.108e-01] | 1.067e-01 [8.489e-02,1.285e-01] | 1.150e-01 [9.989e-02,1.345e-01] | 8.771e-02 [5.595e-02,1.181e-01] | 1.797e-01 [1.201e-01,2.524e-01] |
| ang_acc_max | 1.369e-01 [1.338e-01,1.398e-01] | 5.673e-02 [4.656e-02,6.787e-02] | 3.454e-02 [2.812e-02,4.209e-02] | 1.170e-01 [9.429e-02,1.496e-01] | 1.097e-01 [8.755e-02,1.316e-01] |

### C

| BIC Name | Dataset1 (mean, 95%CI) | Dataset2 (mean, 95%CI) | Dataset3 (mean, 95%CI) | Dataset4 (mean, 95%CI) | Dataset5 (mean, 95%CI) |
|---|---|---|---|---|---|
| SI | 4.165e-01 [4.006e-01,4.324e-01] | 2.607e-01 [2.095e-01,4.224e-01] | 4.222e-02 [2.391e-02,1.007e-01] | 1.161e-01 [7.583e-02,1.879e-01] | 9.794e-02 [7.891e-02,1.321e-01] |
| HIC | 4.159e-01 [3.992e-01,4.319e-01] | 3.145e-01 [2.443e-01,5.125e-01] | 4.385e-02 [2.275e-02,1.039e-01] | 1.368e-01 [8.990e-02,2.153e-01] | 8.586e-02 [6.807e-02,1.173e-01] |
| GAMBIT | 1.910e-01 [1.846e-01,1.970e-01] | 1.648e-01 [1.372e-01,1.879e-01] | 5.907e-02 [4.550e-02,7.179e-02] | 8.957e-02 [5.642e-02,1.377e-01] | 1.429e-01 [1.217e-01,1.616e-01] |
| HIP | 2.718e-01 [2.629e-01,2.803e-01] | 2.908e-01 [2.393e-01,3.525e-01] | 6.251e-02 [3.696e-02,1.132e-01] | 2.899e-02 [1.038e-02,4.478e-02] | 7.894e-02 [6.462e-02,9.662e-02] |
| PCS | 2.707e-01 [2.639e-01,2.774e-01] | 1.687e-01 [1.459e-01,1.902e-01] | 5.328e-02 [3.940e-02,7.705e-02] | 1.570e-01 [1.109e-01,2.297e-01] | 1.262e-01 [1.068e-01,1.515e-01] |
| BRIC | 1.920e-01 [1.875e-01,1.968e-01] | 2.130e-01 [1.552e-01,2.608e-01] | 1.687e-01 [1.519e-01,1.804e-01] | 1.479e-01 [9.004e-02,2.212e-01] | 2.466e-01 [1.888e-01,3.077e-01] |
| PRHIC | 8.346e-01 [7.846e-01,8.849e-01] | 7.854e-01 [4.379e-01,1.763e+00] | 1.125e-01 [2.153e-02,8.816e-01] | 6.020e-01 [2.673e-01,1.310e+00] | 3.819e-01 [9.490e-02,1.198e+00] |
| KLC | 2.407e-01 [2.342e-01,2.474e-01] | 2.463e-01 [2.142e-01,2.846e-01] | 6.308e-02 [3.586e-02,1.279e-01] | 1.492e-01 [1.017e-01,2.075e-01] | 1.179e-01 [9.594e-02,1.509e-01] |
| RIC | 4.419e-01 [4.279e-01,4.569e-01] | 1.649e-01 [1.380e-01,2.179e-01] | 4.424e-02 [3.112e-02,6.441e-02] | 2.024e-01 [1.572e-01,3.399e-01] | 2.573e-01 [2.064e-01,3.200e-01] |
| BrIC | 1.943e-01 [1.902e-01,1.989e-01] | 1.957e-01 [1.375e-01,2.465e-01] | 1.502e-01 [1.145e-01,1.797e-01] | 1.307e-01 [8.379e-02,1.849e-01] | 2.131e-01 [1.490e-01,2.723e-01] |
| CP | 2.202e-01 [2.050e-01,2.354e-01] | 1.155e-01 [9.084e-02,1.358e-01] | 6.147e-02 [4.355e-02,7.949e-02] | 6.306e-02 [3.751e-02,9.251e-02] | 9.932e-02 [8.100e-02,1.154e-01] |
| RVCI | 2.013e-01 [1.984e-01,2.044e-01] | 1.885e-01 [1.666e-01,2.052e-01] | 1.553e-01 [1.276e-01,1.770e-01] | 1.823e-01 [1.436e-01,2.335e-01] | 1.957e-01 [1.591e-01,2.230e-01] |
| CIBIC | 1.986e-01 [1.944e-01,2.031e-01] | 2.001e-01 [1.462e-01,2.469e-01] | 1.635e-01 [1.146e-01,2.086e-01] | 8.895e-02 [6.473e-02,1.195e-01] | 1.972e-01 [1.508e-01,2.355e-01] |
| Damage | 1.997e-01 [1.962e-01,2.035e-01] | 2.085e-01 [1.640e-01,2.437e-01] | 1.840e-01 [1.382e-01,2.245e-01] | 1.165e-01 [9.044e-02,1.528e-01] | 2.161e-01 [1.630e-01,2.612e-01] |
| BAM | 2.264e-01 [2.216e-01,2.312e-01] | 1.283e-01 [1.020e-01,1.510e-01] | 5.702e-02 [4.407e-02,7.450e-02] | 1.688e-01 [1.273e-01,2.244e-01] | 1.411e-01 [1.180e-01,1.635e-01] |
| lin_acc_CG_max | 1.800e-01 [1.734e-01,1.859e-01] | 2.030e-01 [1.693e-01,2.300e-01] | 5.553e-02 [4.145e-02,6.912e-02] | 7.131e-02 [3.959e-02,1.074e-01] | 1.295e-01 [1.094e-01,1.479e-01] |
| ang_vel_max | 1.920e-01 [1.874e-01,1.968e-01] | 2.105e-01 [1.514e-01,2.608e-01] | 1.725e-01 [1.336e-01,2.023e-01] | 1.448e-01 [8.644e-02,1.997e-01] | 2.482e-01 [1.886e-01,3.129e-01] |
| ang_acc_max | 2.260e-01 [2.206e-01,2.311e-01] | 1.310e-01 [1.042e-01,1.537e-01] | 6.171e-02 [4.600e-02,7.492e-02] | 1.722e-01 [1.348e-01,2.213e-01] | 1.414e-01 [1.165e-01,1.650e-01] |

**Supplementary Figure 1** The slopes in the OLS regression of MPS95/MPSCC95/CSDM on each of the 18 BIC on five datasets of head impacts with 1000 iterations of bootstrapping resampling. In the study of each BIC for each response, the BIC is standardized. (A) The slopes in the regression of MPS95. (B) The slopes in the regression of MPSCC95. (C) The slopes in the regression of CSDM.

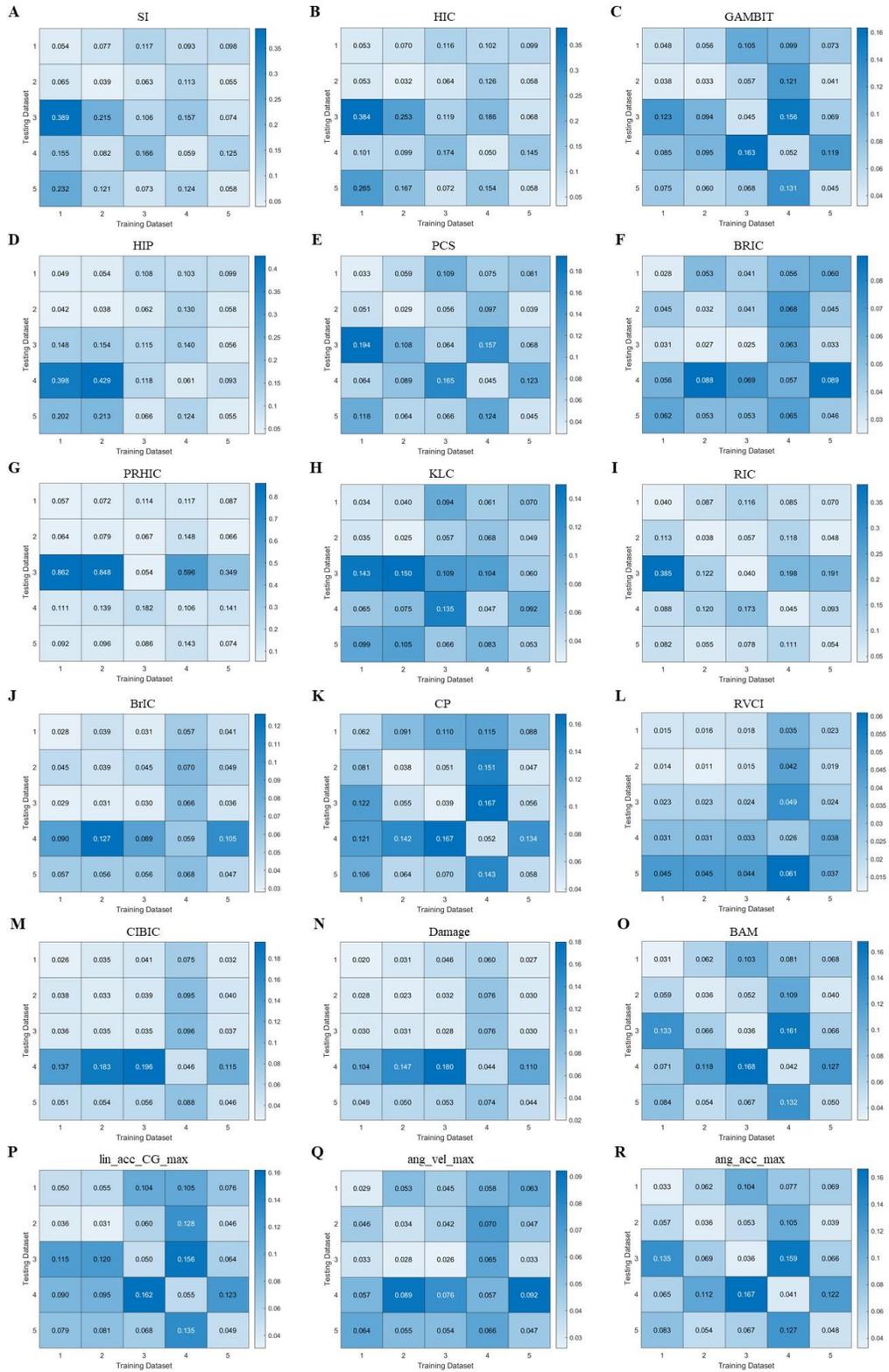

**Supplementary Figure 2** The mean RMSE in the single-dataset prediction and cross-dataset prediction of MPS95 based on 18 BIC with 100 iterations of bootstrapping resampling. Dataset 1: lab impacts; Dataset 2: college football; Dataset 3: MMA; Dataset 4: NHTSA; Dataset5: NASCAR.

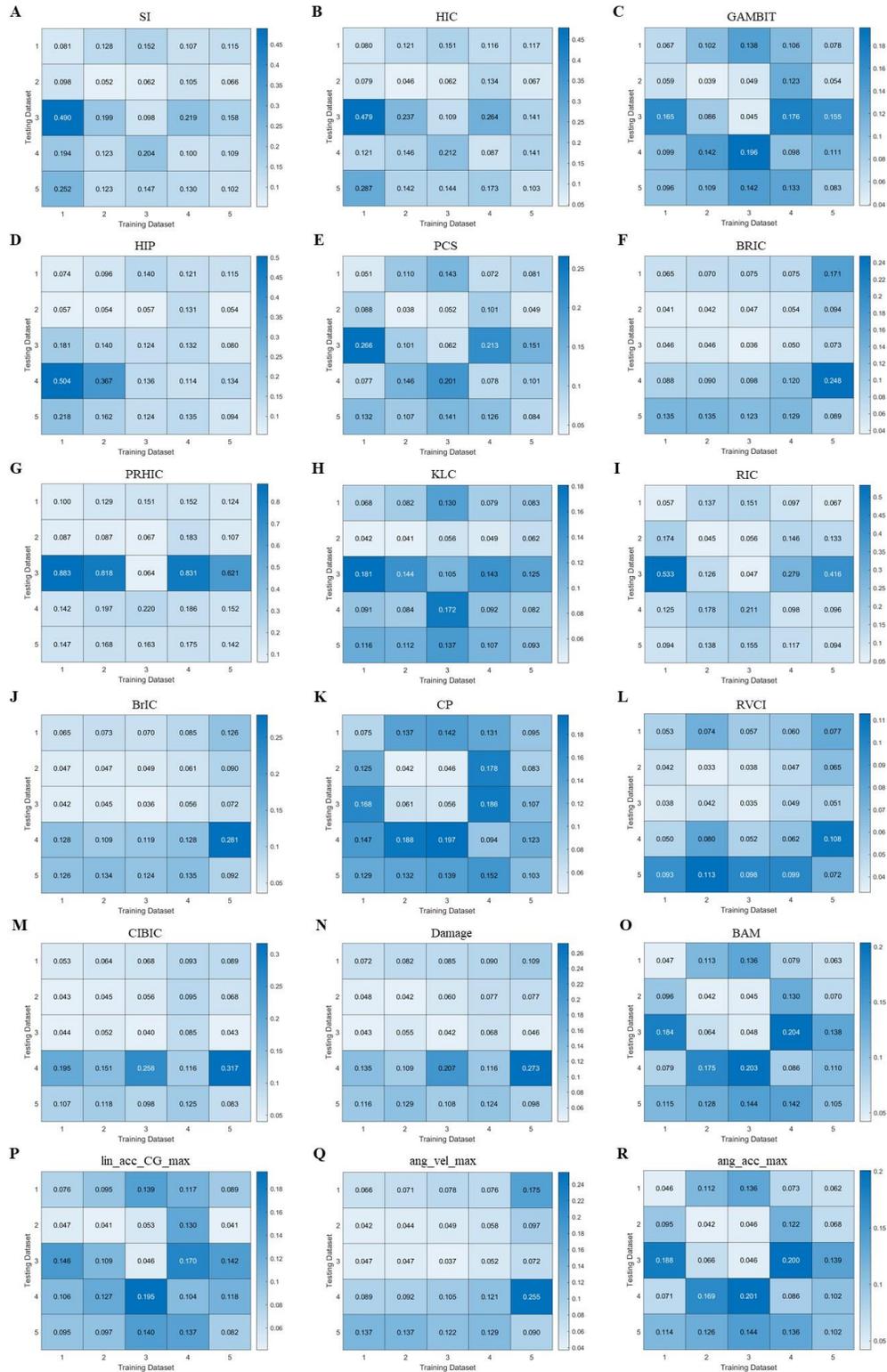

**Supplementary Figure 3** The mean RMSE in the single-dataset prediction and cross-dataset prediction of MPSCC95 based on 18 BIC with 100 iterations of bootstrapping resampling. Dataset 1: lab impacts; Dataset 2: college football; Dataset 3: MMA; Dataset 4: NHTSA; Dataset5: NASCAR.

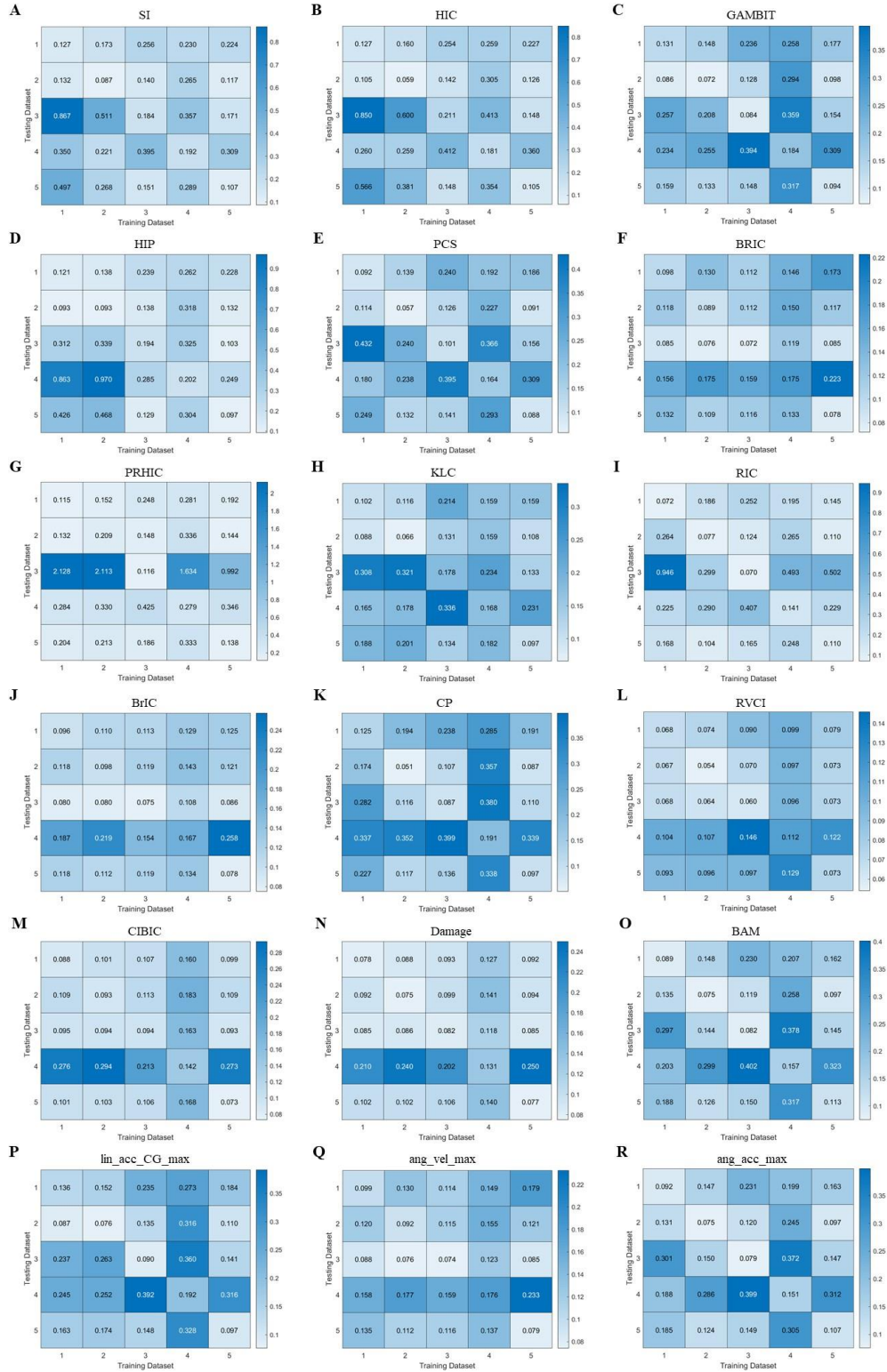

**Supplementary Figure 4** The mean RMSE in the single-dataset prediction and cross-dataset prediction of CSDM based on 18 BIC with 100 iterations of bootstrapping resampling. Dataset 1: lab impacts; Dataset 2: college football; Dataset 3: MMA; Dataset 4: NHTSA; Dataset5: NASCAR.